N. Kayser-Bril
Journalism++,

A. Valeeva
University of Siegen,

I. Radchenko
ITMO University

# TRANSFORMATION OF COMMUNICATION PROCESSES: DATA JOURNALISM

## Introduction

Data journalism can be mistaken as the new buzzword for infographics. This is true, but only to some extent. Visualizing data to tell stories started long before the Internet came into being. What makes an attractive infographics, is not the great design, but, more importantly, the insight it gives. The John Snow's map of cholera outbreaks from nineteenth century London, is an example of reference.

The English physician mapped the cases of cholera deaths in the Soho district of London. Back then, the very notion of germs did not exist. Being mapped, the plain data gave an insight: the outliers clustered around the pump. This is just one of many examples of maps and charts that give another perspective on the same dataset.

Another important part of data journalism – or, more precisely, data-driven journalism, is the use of computers, math and statistical analysis. Computer-assisted reporting was born in the American newsroom in the late 1960s, where technological advances met the social sciences. It was during the US national elections in 1952 when journalists first used a computer to predict the outcome of the vote – and the machine got it right.

Fifteen years later Philip Meyer, a journalist working at the Detroit Free Press, used IBM 360 to cover the Detroit riots in 1967. Through a machine survey, he was able to investigate the dataset and sketch a profile of the rioters. This Pulitzer-winning story was the first attempt by a journalist to use analytical methods from sociology, behavioral science research methods and similar domains in the context of a newsroom. Philip Meyer himself called this "precision journalism."

The increasing size of datasets was another step on the way to data-driven journalism. In 2006, Adrian Holovaty, an American web developer, journalist and entrepreneur, wrote a blogpost which turned out to be a manifesto for data-driven journalism.[1] His main point was that "newspapers need to stop the story-centric worldview." What is required from media, is "structured information: the type of information that can be sliced-and-diced, in an automated fashion, by computers." As Holovaty wrote, the media of the future "have to build an infrastructure that turns them into reliable data hubs, able to analyze even very large and complex datasets internally and to build stories on their insights."

---

[1] See <http://chicago.everyblock.com/crime/ is an example of Holovaty's implementation of his own manifesto>.

What changed in the late 2000's was that Internet gave access to an unprecedented wealth of information and computing power. Data from the public bodies and corporations are becoming increasingly available, in a movement known as open data. The vast amount of information calls for new methods to find and convey meaning from the original data.

From the other end, the tools to handle data permit the evolution in the newsroom to happen. What used to be the exclusive domain of computer scientists can now be done by any journalist. Free software allows anyone to manage, analyze and visualize data. Open source alternatives abound to analyze geographic data or visualize it. Further, solutions on how to scrape, collect and store vast amounts of data are starting to come to the market.

### Newsrooms meet data journalism

The first major news organization to adopt the term is The Guardian, which launched its Datablog in March 2009. Its editor Simon Rogers and his team regularly published stories based on a dataset which he also made public for others to use.

At the same time in France, Le Post (now *Le Huffington Post*), a property of Le Monde, commissioned an interactive ranking of French members of Parliaments (MPs) summarizing the number elective mandates they held besides their position as MPs. The work involved scraping, visualization and investigation and was done by a team of external contractors,[2] in-house developers and journalists. What would today be called data journalism was then done with no clear framework and no descriptive term.

These first experiments created the foundations of the European data journalism for three reasons. The first one is technical. The second is political. The third one is circumstantial.

In 2007, Apple launched the first version of its iPhone. It did not allow Adobe's Flash software to run on the device, mostly as a way to save power. Flash had been the ubiquitous tool for interactive content for the previous ten years. Apple's move meant that content produced with Flash would not be visible on the iPhone and later on the iPad. It fastened dramatically the demise of Flash and the rise of JavaScript-based, browser-rendered interactives. Without delving into technical details, it meant that new skills had to be harnessed as content producers moved away from Flash. These new, non-Flash teams would be the basis for the upcoming data journalism ones.

---

[2] Disclosure: Nicolas Kayser-Bril was one of them.

The second catalyst for the rise of data journalism lies in the dynamism of the open data movement. Outside of the newsrooms, groups of computer developers and activists began to challenge the monopoly of news organizations. The best example is the "Parliament watch" initiatives. Those were websites that measured and structured parliamentary activities. In France, it was an NGO called "Regards Citoyens", which started in 2008 in the wake of the passing of the Hadopi law, which allowed for automated discontinuation of a contract between an internet service provider and its client if the latter was suspected of sharing copyrighted materials. In the late 2000's, Slovakia's Zuzana Wienk created the Fair Play Alliance, which focused on disclosing money flows to and from the public administration by building data-driven online tools. The biggest data-related NGO in Europe however remains the Open Knowledge (OK), which came to continent-wide popularity with the release of the dataset management software CKAN, which was adopted by several open data portals.

New agents also emerged – international journalism groups such as the International Consortium for Investigative Journalists or JournalismFund.eu and specialized agencies like OpenDataCity in Berlin, Journalism++ in five European cities or Dataninja in Bologna.

These NGOs and data journalism teams both considered that development skills should be used to gather and communicate socially relevant information. It is not rare for data journalists to have worked at an NGO committed to open data.[3]

The third reason that explains the rise of journalism at the beginning of the 2010's in Europe lies in Wikileaks' publication strategy for the Afghan Warlogs in June 2010. From 2006 to 2010, Wikileaks, headed by Julian Assange, had published documents only on its own website. To maximize impact, Wikileaks tried a new strategy, setting up a consortium with The Guardian and Der Spiegel in Europe and the New York Times in the United States. For the first time, traditional newsrooms had to deal with source material in database format and needed the skills of working with data. This gave small data teams an opportunity to gain acceptance within their newsrooms. At The Guardian, Simon Rogers and others worked with journalists on the documents. In France, OWNI, a 4-person news operation with a focus for data journalism, worked on the Afghan documents once they were published, helping popularize the concept of data journalism in other newsrooms. Subsequent leaks of databases, organized by Wikileaks and later, by ICIJ (International Consortium of Investigative Journalists), Al Jazeera and others, continued to pressure publishers into paying attention to this trend.

---

[3] For instance, developers Stefan Wehrmeyer and Friedrich Lindenberg both worked for Open Knowledge Germany before or after working in the newsroom of Spiegel or Correct!v and developer-journalist Annabel Church co-organized events in Berlin with OK's Lucy Chambers.

The combination of Adobe Flash's demise, the rapid push of open data and database leaks moved data journalism from a side practice into a widely accepted concept. That data journalism courses started at the same time in journalism schools across Europe is no coincidence.[4]

## Stories from the databases

Databases allow for different modes of work and types of participation. Dealing with a large dataset is complicated if feasible at all for one person alone. Collaboration with others – both inside and outside the newsroom – becomes inevitable. However, it is only for good: it strengthens the journalism culture, restores the trust between the media and the audience and makes more sense of data. There are different types of database-driven journalism stories.

### Shared database for journalists from different media

Media have to engage in a collaborative publishing endeavor: to publish the information on one day and to share the common database. In the journalist spirit of competition and rivalry, this might seem problematic; however this relieves the burden of the data and increases the impact of the publication.

Afghan Warlogs publication was a case when one database was shared between three mainstream media: The Guardian, Der Spiegel and the NYT. The reporters of all three media outlets have joined forces, using a large database to organize the material, then plugging in search terms and combining the documents for newsworthy content. Together they assembled the material into a conveniently searchable and secure database. They brainstormed topics to explore and exchanged search results.

### Global collaboration between journalists

Cross-border investigations allow for a broader reach and localized focus at the same time. The Offshore Leaks project, led by ICIJ, allowed 110 reporters from 47 countries to work together. Gerard Ryle, an Australian journalist, received a hard drive containing 260 gigabytes of information, concerning more than 100.000 companies operating under tax havens. The publication of their findings, on the same day, helped to create a large impact and offer a wider range of coverage on the issue in question. The Offshore Leaks investigation was the largest project of its kind at the time.

---

[4] At Ecole Supérieure de Journalisme, in Lille, France, courses started in 2011.

### Joint forces of journalists, activists and technologists

Confiscati Bene (literally, "Well Confiscated") is a participatory project aiming at stimulating an effective re-use of the assets seized from the mafia.[5] ConfiscatiBene is carried out by a diverse group of journalists, activists and technologists. The project investigates the current condition of the assets through the analysis of relevant data coming both from official sources and from bottom-up, citizen monitoring initiatives.

### Crowdsourcing

Outside collaboration can be done via crowdsourcing. The Guardian chose in 2009 to ask for the help of its readers. The expenses of all the Members of Parliament were published in June 2009 by the House of the Commons.[6] But the amount of files, and the way they were edited led the journalists to build a microsite allowing readers to read and annotate the 700.000 documents. The main goal was to enable "users to fully investigate the documents and track what they — and other users — found", Janine Gibson, editor in chief of The Guardian website, explained in a press release. The Guardian experiment worked well because readers could identify and look for their own Member of Parliament.

### Open Source Intelligence (OSINT)

This method implies collecting already published information to find news. One of the best-known examples is The Migrant Files[7] where a team of ten data-journalists across Europe gathered publicly available information to build an exclusive database of all the migrants and refugees' deaths at the European Union's borders. All the information was available but spread across a variety of data sources from news report to spreadsheets maintained by NGOs. The main task was to gather the information, clean it and construct a single database. On April 2$^d$ 2014, the team published the database and the articles on nine websites across Europe. Their investigation helped to assess the danger of each route taken by the migrants and quantified the deaths not counted by the EU officials. To publish the database, the team

---

[5] See <http://www.confiscatibene.it/it/about-project-english-version>.
[6] See <http://www.theguardian.com/news/datablog/2009/jun/18/mps-expenses-house-ofcommons>.
[7] See <https://www.detective.io/detective/the-migrants-files/>.

used Detective.io, a tool developed by Journalism++ to help journalists make their own databases.

### Database for users to research

After extensive data collection and analysis, data teams can come up with apps or interactive graphics that allow the end user to get insight relevant for him personally. For instance, Dollars for Docs by Propublica[8] allow users to see how their doctors are sponsored by big pharmaceutical companies, and the famous "Rent or Buy[9]" calculator by the New York Times helps to take the right decision on housing.

## Evolution of data journalism after 2010

Newsrooms kept experimenting with data journalism. Zeit Online, in Berlin, commissioned several interactive projects from external developers in 2011. In Paris, Le Monde hired a developer-journalist from OWNI in 2011. All across Europe, in newsrooms small and large, individuals tried tools and techniques. Top management almost always ignored, let alone support, these experiments (a notable exception might be Zeit Online, where executives directly pushed for innovation).

Some newsrooms did capitalize on their employees' efforts at data journalism and set up processes, teams and positions dedicated to the practice. Zeit Online created positions as "developer-journalist" in 2012. In Zurich, Neue Zurcher Zeitung created a dedicated data journalism team that same year with Sylke Grunwald at its helm. In London, the Financial Times reorganized its interactive team under the guidance of Martin Stabe in 2013.

The size of the newsroom is no indicator of its willingness to embrace data journalism. Regional outlets, such as Stimme.de in Heilbronn or L'Avenir, a regional newspapers group in Belgium, have programs to push for data journalism since 2013. Some media outlets that came late in the game did catch up by rapidly creating data teams.

However, some media like Le Monde and Le Figaro, being among the first to experiment with data journalism, did not create teams where developers and journalists could work together. There, developers remain in their own, purely technical teams. As a result, all data journalism activities there took place independently from one another, preventing the organization from capitalizing on its experiences and from gaining productivity.

---

[8] See <https://projects.propublica.org/docdollars/>.
[9] See <http://www.nytimes.com/interactive/2014/upshot/buy-rent-calculator.html?abt=0002&abg=1>.

In Russia, several news organizations or NGOs have experimented with data journalism, but no team has emerged that considered the discipline its main activity. Several databases are available for use, among them the state procurement website (http://zakupki.gov.ru/) and the site run by the state treasure of the Russian Federation (http://bus.gov.ru/). There is also an open data portal run by enthusiasts of open data (http://hubofdata.ru/).[10]

One of the best practitioners of data journalism in Russia so far was the infographics team at RIA news agency. After its dismantling, what is left are the agencies specializing in data visualization like Mercator, and few data journalism projects delivered on the ad-hoc basis, such as those published by Slon.ru.

## Conclusion: data journalism will become the norm

Six years after the term gained acceptance, data journalism remains a new and vaguely defined practice. Rather than being defined by content, inputs or outputs, the consistent definition of data journalism has to do with teamwork and processes. The way developers, project managers and reporters work together allow for producing content in a new, more efficient way. Projects such as Swiss Leaks, Migrants Files or Fifa Files show that these processes are now accepted and replicated as needed. Some newsrooms invested in data journalism, by creating new positions, while others remained on the sidelines.

The term data journalism helped frame the debate regarding how journalism was adapting to the new economic and technical landscape and is now being integrated in the concept of "journalism" as a whole. Content production routinely includes developers and project managers, regardless of whether or not people involved call themselves data journalists. Journalism schools are adding, next to their courses in data journalism, classes in code and statistics to their curriculum,[11] next to their data journalism classes.

What defines data journalism — teamwork across disciplines — will soon become the norm in content production. Newsrooms that inherit a print-based brand still have a long way to go before they produce content by having personnel with different skill sets working together as a matter of routine. But as competition on content increases, with NGOs, corporations and institutions joining in the fray, the share of content produced in a single-person, text-only way, will be marginalized.

---

[10] For the full list of Russian portals with data, check <http://iradche.ru/2014/08/open-data-2/>.

[11] Coding classes started in 2013 at the School of Journalism of Sciences-Po Paris, France. Statistics were introduced in 2014 at Académie du Journalisme et des Médias in Neuchâtel, Switzerland.

The digitalization of content has radically transformed the media industry. Now that every Internet user can commit acts of journalism, professional journalists are in dire need of a redefinition of their work and purpose. With the rise of data, journalists can move up the value chain and focus on collecting information and analyzing it. Therefore, they can use the data as a resource for social change.

## References

*Gray J., Bounegru L., Chambers L*. Data Journalism Handbook // O'Reilly. 2012.

*Holovaty A*. A Fundamental Way Newspaper Sites Need to Change, 2006 <http://www.holovaty.com/writing/fundamental-change/>.

*Howard A*. The Growing Importance of Data Journalism // O'Reilly. 2010. 21 December.

*Keller B*. Dealing with Assange and the WikiLeaks Secrets // The New York Times. 2011. 26 January <http://www.nytimes.com/2011/01/30/magazine/30Wikileaks-t.html?pagewanted=all&_r=1&>.

*Léchenet A*. Global Database Investigations: The Role of the Computer-Assisted Reporter / Reuters Institute Fellowship Paper. University of Oxford, 2014.

*Lorenz M*. From Attention to Trust: Data-Driven Journalism and the Urban Future. Presentation // Picnic. Amsterdam, Netherlands, 2011. 23 September.

*Lorenz M., Kayser-Bril N., McGhee G*. Voices: News Organizations Must Become Hubs of Trusted Data in a Market Seeking (and Valuing) Trust // Nieman Lab. 2011. 1 March.

*Rogers S*. Facts are Sacred. Faber and Faber, 2013.